# Germanium crystals on silicon show their light

F. Pezzoli[1*], F. Isa[2], G. Isella[2], C.V. Falub[3], T. Kreiliger[3], M. Salvalaglio[1], R. Bergamaschini[1], E. Grilli[1], M. Guzzi[1], H. von Känel[3] and Leo Miglio[1]

[1] LNESS and Dipartimento di Scienza dei Materiali, Università degli Studi di Milano-Bicocca, Via R. Cozzi, I-20125 Milan, Italy.

[2] LNESS and Dipartimento di Fisica, Politecnico di Milano, Piazza Leonardo da Vinci 32, I-20133 Milan, Italy.

[3] Laboratory for Solid State Physics, ETH Zürich, Schafmattstrasse 16, CH-8093 Zürich, Switzerland

\* Corresponding author, fabio.pezzoli@unimib.it

**Germanium and silicon-germanium alloys have found entry into Si technology thanks to their compatibility with Si processing and their ability to tailor electronic properties by strain and band-gap engineering. Germanium's potential to extend Si functionalities, as exemplified by lasing action of strained-Ge on Si substrates, has brought the material back to attention. Yet despite these advances, non-radiative transitions, induced by crystal defects originating from the Ge/Si interface, continue to be a serious bottleneck. Here we demonstrate the drastic emission enhancement achieved via control and mitigation over the parasitic activity of defects in microscale Ge/Si crystals. We unravel how defects affect interband luminescence and minimize their influence by controlling carrier diffusion with band-gap-engineered reflectors. We finally extended this approach designing efficient quantum well emitters. Our results pave the way for the large-scale implementation of advanced electronic and photonic structures unaffected by the ubiquitous presence of defects developed at epitaxial interfaces.**

Silicon offers an ideal platform for hybrid[1-2] and monolithic architectures[3-8] exploiting novel electronic and photonics functionalities[9]. While hybrid solutions may be less demanding in terms of materials compatibility, they tend to be rather ill suited for large-scale integration[10]. Monolithic integration on the other hand, while allowing processing at a wafer scale, faces well-known defect problems characteristic of heteroepitaxial growth[11,12]. Here, we consider the Ge/Si(001) system. Ge has a high light-matter interaction due to its band structure with a quasi-direct band gap[3,11] in many respects mimicking the behaviour of III-V semiconductors. As opposed to these, Ge shares the same diamond crystal structure with the Si substrate and it is compatible with Si processing, which greatly simplifies monolithic integration schemes[11-13]. Its optical response can be tuned between the E and L communication bands[7,8,12,14] and it offers outstanding hole transport properties[15]. Nevertheless, as happens to other materials systems with significantly different lattice parameters, heteroepitaxial growth of Ge on Si is prone to defect formation in the form of misfit dislocations at the substrate-film interface and nucleated to release the strain imposed by initially bringing the two crystal lattices into registry[16,17]. If misfit dislocations were the only culprits spoiling device performance, their influence could easily be removed by placing devices far from the interface. Unfortunately, misfit dislocations are always accompanied by threading arms extending to the free surface and hence inevitably piercing through any active layer[17]. Indeed, a wealth of applications based upon low-defect Ge have been deferred so far, since the various approaches, proposed for reducing the threading dislocation densities (TDD)[18-25], yet resulted in TDD values far above those attainable by bulk-Ge crystal growth[17,26]. In this work we demonstrate the remarkable suppression of the unwanted, non-radiative activity of dislocations. The reliable and effective control over defects, offered by epitaxial seeded growth[27], manifests itself in the Ge on Si system by a surge of the interband (IB) photoluminescence intensity by more than two orders of magnitude with respect to that typical for epitaxial layers directly grown on planar substrates. We deepen the understanding of the optical properties of the Ge/Si system by taking full advantage of bandgap engineering to identify the location of non-radiative recombination centers and to study the interplay of IB and dislocation-related emission. This allows us to glimpse the carrier diffusion and recombination mechanisms and get insights into the drastic emission enhancement in Ge QW emitters realized on top of the defect-free crystals. Our findings have the potential to fertilize numerous applications of Ge, such as advanced electronics with functional materials, hot hole Ge lasers for lab-on-chip terahertz spectroscopy and near-infrared imaging detectors for sensing biological targets.

## Results

**Ge on Si(001) heteroepitaxy.** Figure 1 shows the morphology of typical 8-μm-tall Ge-crystals epitaxially grown[28] onto 2×2 μm$^2$, 5×5 μm$^2$ and 9×9 μm$^2$ Si(001) pillars. By changing the size of the Si pillars we are able to modify the top morphology of the three-dimensional (3D) Ge-crystals, while keeping fixed the growth parameters[27,28]. The deposition on 2×2 μm$^2$ Si pillars results in Ge-crystals with a pyramidal top surface bounded by {113} and {111} facets, the former being dominant (Fig. 1a). Increasing the size of the Si pillar base to 5 μm leads to a (001) facet in the centre the Ge-crystals (Fig. 1 b). When the width of the Si pillar is further increased, the (001) facet expands so that the crystals grown on the 9×9 μm$^2$ mesa have almost flat (001) tops (Fig 1c). An overview of the top surface coverage by the (001) facet is given in Fig. 1d.

Etch pit count revealed that the uniform Ge(001) layer grown during the same deposition run on the unpatterned region of the Si substrate is affected by threading dislocations with a density of $(3.5\pm0.9)\times10^8$ cm$^{-2}$. The facet structure of the Ge-crystals has a strong impact on the distribution of threading dislocations, since dislocation arms are deflected towards the sidewalls by the inclined growth front[27,29]. As a result, in fully {113} faceted crystals, dislocation arms are confined to the very first microns close to the Ge/Si interface, while the upper part remains dislocation-free[27]: a noteworthy result achieved at a relatively low deposition temperature, i.e. 550°C, and without additional thermal treatments. Since the mechanism of dislocation deflection is ineffective for the (001) top-facet, dislocations running along the growth direction will remain trapped underneath the (001) surface right at the centre of the Ge-crystals. The average dislocation density approaches that found in continuous films by increasing the size of the Ge-crystals, i.e. for a larger (001) surface coverage.

**Photoluminescence experiments on Ge-crystals.** Here, we address the impact of crystal morphology and concomitant dislocation density on the light emission efficiency. Figure 2a shows room temperature (RT) photoluminescence (PL) spectra of Ge crystal-arrays grown on patterned Si substrates with 2×2 μm$^2$ (blue line), 5×5 μm$^2$ (red line), 9×9 μm$^2$ (orange line) pillars. For comparison, the corresponding spectrum recorded on the unpatterned area of the same sample is also shown (black line). The measurement conditions are identical for all the samples.
The large number of defects in the continuous Ge film present in the unpatterned substrate areas drastically affects the overall PL efficiency. There, carrier dynamics is dictated by the parasitic, non-radiative channels. The time scale for these processes is much shorter than that of the radiative

IB transitions, such that the luminescence intensity falls well below the measurement noise (black line in Fig. 2a). This is in agreement with reports on typical undoped Ge/Si heteroepitaxial layers, where no PL signal could be detected in as-grown samples[30]. By contrast, Fig. 2a reveals a rapid increase of the PL intensity for as-grown 3D Ge-crystals with decreasing size. Fully faceted Ge-crystals grown on 2×2 μm² Si pillars yield the highest emission intensity: at least a factor of ~300 above the noise level of the corresponding continuous film. This value is even underestimated, since it does not take into account the smaller surface coverage of the crystal arrays and its dependence on the pattern size (see Methods). Such remarkable result is achieved without the need of post-growth annealing and no surface treatment to ensure a low recombination velocity, despite the large surface-to-volume ratio of the 3D crystals.

According to Fig. 2b the PL of the Ge-crystals (blue line) consists of two spectral features. The main PL structure has a maximum at ~0.7 eV, i.e. close to the energy of the fundamental gap. It is due to recombination of electrons in the lowest indirect conduction band minimum at the L point of the Brillouin zone with holes at the top of the valence band at k = 0, $L_c - \Gamma_v$ [30,31]. The shoulder at ~0.77 eV is due to recombination across the direct $\Gamma_c-\Gamma_v$ gap. In Fig. 2b we compare the PL obtained from our epitaxial, nominally undoped Ge-crystals with that of an n-type Ge-wafer (gray line), with an As content of ~8.3×10$^{16}$ cm$^{-3}$.

The PL spectrum of epitaxial Ge-crystals has a spectral shape identical to that of the Ge-wafer. It is characterized by a weaker amplitude $I^d$ of the direct transition with respect to that of the indirect transition $I^i$. By contrast, all the previous literature reports about Ge/Si heterostructures[30-37] pointed out that the direct gap dominates over the indirect gap emission, so that $I^d > I^i$. This may be understood by considering that a high density of defects, besides reducing the overall PL efficiency, can affect carrier distribution in the conduction band and shift the spectral weight of the emitted light in favour of the direct gap transition[30], thus leading to $I^d > I^i$. Noteworthy, Ge/Si CVD layers with thicknesses comparable to the one used here display $I^d > I^i$ along with defect-related features in the RT PL spectra[35]. By contrast, in our Ge-crystals, dislocations are confined to spatial regions close to the interface, so that populations of Γ- and L-valleys are not perturbed within the crystal volume. This results in the spectral shape characteristic for a defect-free bulk material and reported in Fig. 2.

**Temperature dependence of interband and dislocation emissions.** Although the PL of epitaxial Ge-crystals carries the fingerprints of defect-free bulk material, the dislocations present in the first few microns close to the interface have an effect on the overall emission efficiency. In order to shed light on the influence of dislocations on the carrier population in the bands, we have performed temperature-dependent PL measurements on Ge-crystals bounded by {113} facets. The wide spectral range covered by the measurements allows us to extend the optical investigations well below the low-energy edge of the Ge emission and to study the interplay between IB recombination and dislocation-related PL (DPL). As shown in Fig. 3a, at 6K the PL spectrum is characterised by direct gap recombination at 0.882 eV, indirect gap recombination assisted by longitudinal acoustical phonon emission at 0.712 eV and the no-phonon line at ~0.736 eV [38]. The additional band, extending from 0.45 eV to about 0.67 eV, is attributed to recombination taking place at dislocations[39]. The broad dislocation spectrum may be possibly due to dissociation of dislocations into Shockley partials and their interaction with point defects[40-42].

In Fig. 3a the color-coded map of the PL intensity shows that the spectra features become broader and weaker with increasing temperature. The direct gap emission and the low energy edge of the indirect gap emission follow the temperature dependence of the direct and indirect gap, respectively[43]. With increasing temperature, the high energy side of the indirect gap PL becomes broader and its amplitude affected by the activation of transitions mediated by phonon absorption[38].

Figure 3b shows the integrated PL intensity of dislocation and IB emission (direct plus indirect gap transitions) versus temperature. Interestingly, the IB luminescence exhibits two temperature regimes. Below a critical temperature, $T_C \sim 125$ K, the PL increases gradually, whereas above $T_C$ it rises sharply. As a matter of fact, for $T>T_C$ the abrupt DPL quenching occurs in phase with the steep increase of the IB emission.

In order to obtain a more detailed understanding of the observed trends, we employ the following charge-dependent potential barrier model[44,45]. Under steady state conditions and for low excitation levels, expressions for the net concentrations $\delta m$ of holes trapped at dislocations, for the excess concentration $\delta n$ of electrons in the conduction band and for the hole concentration $\delta p$ in the valence band, can be derived from rate equations involving a potential barrier between charged line defects and the bulk matrix[45] (see Supplementary Information). At a temperature T, trapped carriers can either be released and contribute to the radiative IB recombination or contribute to the DPL emission or recombine non radiatively at the dislocation site. The inset of Fig. 3b illustrates the physical processes involved. As shown by the solid blue line of Fig. 3, the IB transition is well-described by applying the constraint of crystal neutrality[45], $\delta n = \delta p + \delta m$. According to the model, at low temperatures $\delta p \ll \delta m$, since carriers leak out from the band into the trapping defect levels, thereby enhancing the defect emission. When the temperature is increased, the traps empty by thermal activation such that $\delta p \gg \delta m$. In other words, the IB recombination regime takes over and the IB luminescence becomes dominant. Indeed, as revealed by the experimental data of Fig. 3b, the temperature $T_c$ at which the crossing between the recombination and trapping regimes takes place coincides with the temperature at which $\delta p = \delta m$ and the IB integrated intensity has a minimum[45].

**Interband enhancement via band gap engineering.** Having unravelled the optical activity of dislocations present in the region closest to the buried interface of epitaxial 3D Ge-crystals, we use bandgap engineering to minimize their role. At first we consider coherent SiGe alloy barriers acting as electron and hole reflectors into the Ge-crystals.

Based on the band alignments and effective masses taken from Ref.[46], we have calculated the transmission coefficients for electrons and holes through $Si_{1-x}Ge_x$ barriers within the transfer matrix formalism[47]. Taking as an ansatz a sample at room temperature, with an optically excited carrier density of $1\times10^{19} cm^{-3}$ on the upper side of the barrier, we then derive the fraction of carriers crossing the barrier towards the dislocated region. Figure 4a shows the transmitted fraction of carriers versus the Ge content, x, for a 10 nm wide $Si_{1-x}Ge_x$ reflector. The fraction of transmitted holes increases monotonically with x as a result of the decreasing valence band edge offset (see Supplementary Information). Surprisingly, the transmitted fraction of electrons first decreases with increasing x, and then slowly increases again. As discussed in detail in the Supplementary Information, this counter-intuitive prediction can be rationalised as a result of the cross-over between the $\Delta$ and L valleys of the SiGe band structure under tensile strain.

Since, as summarized in Fig. 4a, the alloy concentration required for optimum performance of these reflectors is around the minimum of the transmission probability for electrons, we have grown three samples with Ge-crystals on 2-μm-wide Si pillars with 10 nm thick $Si_{0.25}Ge_{0.75}$ barriers introduced at a distance of 6 μm (A), 4 μm (B), and 2 μm (C) from the dislocated Si/Ge interface. Finally, a fourth sample contains all three barriers. It should be noted that the thickness of the $Si_{0.25}Ge_{0.75}$ reflectors is optimized so that plastic strain relaxation is avoided.

Figure 4b shows the IB integrated intensity of the Ge-crystals. At low temperature and in the trapping regime, the spectral weight of the PL bands is modulated by the competition of capturing the carriers between the optical recombination centres and traps. Figure 4b clearly demonstrates that the SiGe reflectors improve the radiative transitions across the Ge gaps, confirming that dislocations in the interfacial region are responsible for the PL quenching of the IB emission. The rate of recombination at dislocations and the role of SiGe reflectors is further clarified by calculations of carrier diffusion summarized in Fig 4c (see also Supplementary Information). As the reflector is pushed towards the dislocated interface, the number of carriers, N, within the overall volume of the Ge-crystal increases, thus incrementing the IB recombination events. Noteworthy, the largest improvement of the IB/DPL ratio is achieved for the sample with three embedded SiGe reflectors being this configuration more efficient in suppressing carrier depletion by the buried defects.

**Optical properties of quantum well crystals.** We pushed forward our approach aiming at the full confinement of carriers in the defect free volume of the crystals, i.e. away from the dislocated buried interface. To demonstrate this concept, we employ a larger number of SiGe reflectors, designing Ge quantum wells (QW) embedded in Ge-rich SiGe barriers. Such structures are known to possess a type-I band alignment, which confines holes inside the Ge well and is effective in confining both L- and Γ- valleys electrons[5,48,50].

The sample investigated here features 50 strain-compensated QWs deposited on top of a 8-μm-thick SiGe buffer grown on a 2×2 μm$^2$ Si mesa. The SEM micrograph reported in Fig 5a-b shows that the QW-crystals are bounded by {113} facets, and that the typical morphology of fully faceted 3D crystals is not affected by the presence of the heterostructures. The high-resolution X-ray diffraction measurements, shown in Fig 5c, confirm that the SiGe buffer is fully relaxed[27] and that the QW structure is lattice-matched to the top of the buffer, which contains (91±1)% of Ge.

In the (004) reciprocal space map of Fig 5c, the coherent Ge QWs grown on high index {113} facets give rise to diffraction maxima (red arrows) along the dashed [001] and [113] lines. The structural parameters of the QW stack in the 3D crystals are deduced from the spacing of these diffractions peaks. The Ge content of the barrier is (86 ±1)%, while the thickness of the QW and barrier is of about 15 and 23 nm, respectively. A second kind of satellites, depicted by green arrows along the [001] line in Fig. 5c, arise from the thermally strained QW present in the trenches between neighbouring pillars.

The huge impact of 3D heteropitaxial growth on the optical properties is disclosed by the PL spectra shown in Fig 5d, where the low temperature luminescence of QW-crystals is compared with the negligible PL of a reference layer deposited on the flat Si(001) substrate during the same growth. Remarkably, DPL dominates the spectrum measured on the unpatterned areas but it is strongly suppressed in the QW-crystals. Here quantum-confined transitions related to the direct gap at the Γ point and the indirect gap at the L point of the Ge QWs are clearly visible. We tentatively attribute the PL peak at 0.959 eV to the fundamental transition between the confined state at the Γ point in the conduction band and the heavy hole subband, namely cΓ1-HH1. The low-energy PL doublet is ascribed to dipole allowed transitions across the indirect gap. The high-energy part of the doublet at 0.745 eV is possibly due to the cL1-HH1 recombination of confined carriers, while the line at 0.722 eV is the phonon-assisted optical transition. Such values are in agreement with transition energies reported in the literature for (001)Ge-QWs grown with the same method and similar deposition parameters[50]. Such a benchmark and the lack of DPL due to the effective localization of carriers in the defect-free part of the crystal prove that QW-crystals can outperform high-quality (001)Ge-QWs grown on graded SiGe buffers. This points out the feasibility of monolithic integration of efficient light emitters on silicon, circumventing all the issues associated to the use of thick graded buffers[12,24-27]. We finally note that the growth on the high index facets of 3D crystals might also be beneficial for the heteroepitaxy on Si of active layers based on polar, direct-gap semiconductors[13].

**Discussion**

The Ge/Si system is an excellent model to unfold the physics underlining the various recombination mechanisms effective in heteroepitaxial architectures. The dislocation-related emission, noticeably present in Ge/Si, allows us to underpin the role played by defects in limiting the radiative emission, and to identify viable approaches based upon bandgap engineering to enhance the optical response. Our unique results show that the optical properties of Ge-crystals epitaxially grown onto patterned Si(001) substrates become indistinguishable to those of defect-free Ge-wafers, when carriers are prevented from approaching the dislocated interface. This holds true even for growth temperatures compatible with CMOS-processed substrates, for which the influence of defects on the properties of conventional layered systems is overwhelming. We expect defect-free Ge on Si crystals to greatly facilitate the monolithic integration of advanced devices on large Si-substrates. We anticipate here QW emitters, but an array of applications are already at the grasp of the proposed approach. Defect-free mesa can serve as a platform for introducing novel channel materials capable of extending Si-based CMOS logic, photovoltaics based on 3D crystals can face the long-sought problem of multiple junction solar cells monolithically grown on Si, and microcrystal optical resonators can be applied to next generation photonics links. Indeed, the results outlined in this work for Ge/Si can be extended to other hetero-systems, and they point towards the feasibility of the long-sought integration of diverse materials on the well-established Si platform.

## Methods

***Sample growth.*** Two-dimensional arrays of 2×2, 5×5 and 9×9 $\mu m^2$ pillars with edges along [110] directions were patterned into Si(001) substrates by optical lithography followed by deep reactive ion etching. The pillars were separated by 8 μm deep and 3 μm wide trenches. The patterned wafers were cleaned by standard RCA procedure and then dipped into hydrofluoric acid solution before being transferred into the growth reactor. After the in-situ H desorption, a 100 nm thick Ge buffer was deposited at 500°C at a rate of 0.27 nm/s by low-energy plasma-enhanced chemical vapour deposition (LEPECVD). The temperature was then ramped up to 550°C for the subsequent deposition of 7.9 μm of Ge at a rate of 4.2 nm/s. Immediately after the growth, samples were cooled to room temperature at a rate of ~1°C/s. The surface filling factors are 77%, 89% and 90% for Ge-crystals grown on 2×2 $\mu m^2$, 5×5 $\mu m^2$ and 9×9 $\mu m^2$ Si pillars, respectively.

Samples incorporating 10 nm thick $Si_{0.25}Ge_{0.75}$ reflectors at either 2, or 4, or 6 μm from the Ge/Si interface, as well as at all these three heights were grown according to the aforementioned growth protocol on arrays of 2×2 $\mu m^2$ pillars.

The quantum well (QW) crystal was grown on 2×2 μm$^2$ Si mesa. The first part of the heterostructure was a 8μm-thick $Si_{0.1}Ge_{0.9}$ buffer layer deposited at 625°C at a rate of ~5 nm/s. The multiple QW structure was deposited at 475°C and it consisted of 50 Ge QWs embedded in $Si_{0.15}Ge_{0.85}$ barriers.

***Structural and electrical characterization.*** The deposited material is intentionally undoped, and the residual background doping of the sample with no SiGe barriers was determined by room temperature Hall measurements to be p-type and of ~$(4.2\pm0.8)\times10^{15}$ cm$^{-3}$. All the samples are grown one after the other to ensure the same residual doping level.

High-Resolution X-Ray Diffraction (HRXRD) measurements were performed with Cu K$\alpha_1$ radiation using a diffractometer equipped with a 4-bounce Ge(220) crystal monochromator on the incident beam, an analyzer crystal and a Xe point detector on the diffracted beam. The size of the X-ray beam was ~1 mm.

Threading dislocation densities were obtained by defect etching. The samples were etched in a diluted iodine solution (15 mg $I_2$, 33 ml glacial acetic acid, 10 ml 65% nitric acid, 5 ml 40% HF, etch rate about ~10 nm/s) at 0°C. Etch pits counting was performed by means of Atomic Force Microscopy.

Finally, the morphology of the Ge-crystals was investigated by means of scanning electron microscopy (SEM) using a Zeiss ULTRA 55 digital field emission microscope.

***Optical measurements.*** Optical investigation of the samples was carried out by photoluminescence measurements performed at a variable temperature in a closed-cycle cryostat using a Fourier transform spectrometer, equipped with a Peltier cooled PbS detector. A continuous wave Nd-YVO$_4$ laser operating at 1.165 eV was used as excitation source. At this photon energy, the penetration depth is about 610 nm. The laser spot size on the sample surface had a diameter of ~100 μm, and the corresponding power density was between 1 and 3 kW/cm$^2$.

**References**

1       Fang, A. W. et al. Electrically pumped hybrid AlGaInAs-silicon evanescent laser. *Optics Express* **14,** 9203–9210 (2006).

2       Ashley, T. et al. Heterogeneous InSb quantum well transistors on silicon for ultra-high speed, low power logic applications, *Electronics Letters* **43**, 777-779 (2007).

3       Liang, D. & Bowers, J. E. Recent progress in lasers on silicon. *Nature Photonics* **4**, 511-517 (2010).

4       Liu, J., Sun, X., Camacho-Aguilera, R., Kimerling, L. C. & Michel, J. Ge-on-Si laser operating at room temperature. *Optics Letters* **35,** 679–681 (2010).

5       Pezzoli, F. et al. Optical Spin Injection and Spin Lifetime in Ge Heterostructures. *Physical Review Letters* **108,** 156603 (2012).

6       Cazzanelli, M. et al. Second-harmonic generation in silicon waveguides strained by silicon nitride *Nature Materials* **11**, 148–154 (2012)

7       Soref, R. Mid-infrared photonics in silicon and germanium. *Nature Photonics* **4**, 495-497 (2010).

8       Michel, J., Liu, J. F. & Kimerling, L. C. High-performance Ge-on-Si photodetectors. *Nature Photonics* **4**, 527-534 (2010).

9       Chen, X., Li, C. & Tsang H. K. Device engineering for silicon photonics. *NPG Asia Mater.* **3**, 34–40 (2011).

10      Liang, D., Bowers, J. E. Photonic integration: Si or InP substrates?. *Electronics Letters* **45**, 578-580 (2009)

11      Soref, R. A. SILICON-BASED OPTOELECTRONICS. *Proceedings of the IEEE* **81**, 1687-1706 (1993).

12      Lourdudoss, S. Heteroepitaxy and selective area heteroepitaxy for silicon photonics. *Current Opinion in Solid State & Materials Science* **16**, 2 (2012).

13      Fang, S. F., et al. GALLIUM-ARSENIDE AND OTHER COMPOUND SEMICONDUCTORS ON SILICON. *Journal of Applied Physics* **68**, R31 (1990).

14      Jain, J.R., et al. A micromachining-based technology for enhancing germanium light emission via tensile strain. *Nature Photonics* **6**, 398 (2012).

15      Pillarisetty, R. Academic and industry research progress in germanium nanodevices. *Nature* **479**, 7373 (2011).

16      Matthews, J. W. & Blakeslee, A. E. DEFECTS IN EPITAXIAL MULTILAYERS.1. MISFIT DISLOCATIONS. *Journal of Crystal Growth* **27**, 118-125 (1974).

17      Claeys, C. & Simoen, E., *Extended defects in germanium* (Springer-Verlag, Berlin Heidelberg, 2009).

18      Baribeau, J. M., Jackman, T. E., Houghton, D. C., Maigné, P. and Denhoff, M. W. Growth and characterization of $Si_{1-x}Ge_x$ and Ge epilayers on (100) Si. *Journal of Applied Physics* **63**, 5738 (1988).

19      Fitzgerald, E. A., et al. Nucleation mechanisms and the elimination of misfit dislocations at mismatched interfaces by reduction in growth area. *Journal of Applied Physics Letters* **65**, 2220 (1989).

20      Fitzgerald, E. A., et al. Elimination of interface defects in mismatched epilayers by a reduction in growth area. *Applied Physics Letters* **52**, 1496 (1988).

21      Luan H.C., et al. High-quality Ge epilayers on Si with low threading-dislocation densities. *Applied Physics Letters* **75**, 2909 (1999).

22      Leitz C. W., et al. Dislocation glide and blocking kinetics in compositionally graded SiGe/Si. *Journal of Applied Physics* **90**, 2730 (2001).

23      Loh T.H., et al. Ultrathin low temperature SiGe buffer for the growth of high quality Ge epilayer on Si(100) by ultrahigh vacuum chemical vapor deposition. *Applied Physics Letters* **90**, 092108 (2007).

24      Ayers, J.E. Compliant substrates for heteroepitaxial semiconductor devices: Theory, experiment, and current directions. *J. Electron. Mater* 37, 1511 (2008).

25      Capellini, G., et al. Strain relaxation in high Ge content SiGe layers deposited on Si. *Journal of Applied Physics* **107**, 063504 (2010).

26      Yamamoto Y., et al. Low threading dislocation Ge on Si by combining deposition and etching. *Thin Solid Films* **520**, 3216 (2012).

27      Falub, C. V., et al. Scaling Hetero-Epitaxy from Layers to Three-Dimensional Crystals. *Science* **335**, 1330-1334 (2012).

28      Rosenblad, C., et al. Silicon epitaxy by low-energy plasma enhanced chemical vapor deposition. *Journal of Vacuum Science & Technology a-Vacuum Surfaces and Films* **16**, 2785-2790, (1998).

29      Bai, J., et al. Study of the defect elimination mechanisms in aspect ratio trapping Ge growth. *Applied Physics Letters* **90**, 101902 (2007).

30      Grzybowski, G., et al. Direct versus indirect optical recombination in Ge films grown on Si substrates. *Physical Review B* **84**, 205307 (2011).

31      Kittler, M., et al. Photoluminescence study of Ge containing crystal defects. *Physica Status Solidi A-Applications and Materials Science* **208**, 754–759 (2011).

32      El Kurdi M., et al. Two-dimensional photonic crystals with pure germanium-on-insulator. *Optics Commununications* **281**, 846–850 (2008).

33      Shambat, G., Cheng, S.-L., Lu, J., Nishi, Y. & Vuckovic J. Direct band Ge photoluminescence near 1.6 μm coupled to Ge-on-Si microdisk resonators. *Applied Physics Letters* **97**, 241102 (2010) 34      Jan, S. R., et al. Influence of defects and interface on radiative transition of Ge. *Applied Physics Letters* **98**, 141105 (2011).

35      Arguirov, T., Kittler, M. & Abrosimov, N. V. Room temperature luminescence from Germanium. *Journal of Physics: Conference Series* **281**, 012021 (2011).

36      Nam, D., et al. Electroluminescence from strained germanium membranes and implications for an efficient Si-compatible laser. *Applied Physics Letters* **100**, 131112 (2012).

37      Carroll, L., et al. Direct-Gap Gain and Optical Absorption in Germanium Correlated to the Density of Photoexcited Carriers, Doping, and Strain. *Physical Review Letters* **109**, 057402 (2012).

38      Lieten, R. R., et al. Photoluminescence of bulk germanium. *Physical Review B* **86**, 035204 (2012).

39      Kolyubakin, A. I., Osipyan, Y. A., Shevchenko, S. A. & Shteinman, E. A. DISLOCATION LUMINESCENCE IN GE. *Fizika Tverdogo Tela* **26,** 407 (1984).

40      Shevehenko, S. A. & Tereshchenko, A. N. Photoluminescence in germanium with a quasi-equilibrium dislocation structure. *Physics of the Solid State* **49**, 28–33 (2007).

41      Izotov, A. N., Kolyubakin, A. I., Shevchenko, S. A. & Steinman, E. A. PHOTOLUMINESCENCE AND SPLITTING OF DISLOCATIONS IN GERMANIUM. *Physica Status Solidi a-Applied Research* **130**, 193 (1992).

42      Sauer, R., Kisielowskikemmerich, C. & Alexander, H. DISSOCIATION-WIDTH DEPENDENT RADIATIVE RECOMBINATION OF ELECTRONS AND HOLES AT WIDELY SPLIT DISLOCATIONS IN SILICON. *Physical Review Letters* **57**, 1472-1475 (1986).

43      Varshni, Y. P. Temperature dependence of the energy gap in semiconductors. *Physica* **34**, 149 (1967).

44      Morrison, S. R. Recombination of Electrons and Holes at Dislocations. *Physical Review* **104**, 619-623 (1956).

45      Figielski, T. RECOMBINATION AT DISLOCATIONS. *Solid-State Electronics* **21**, 1403-1412 (1978).

46      Rieger, M. M. & Vogl, P. ELECTRONIC-BAND PARAMETERS IN STRAINED SI(1-X)GE(X) ALLOYS ON SI(1-Y)GE(Y) SUBSTRATES. *Physical Review B* **48**, 14276–14287 (1993).

47      Tsu, R. & Esaki, L. TUNNELING IN A FINITE SUPERLATTICE. *Applied Physics Letters* **22**, 562 (1973).

48      Virgilio, M. & Grosso, G. Type-I alignment and direct fundamental gap in SiGe based heterostructures. *J. Phys.: Condens. Matter* **18**, 1021–1031 (2006) .

49      Claussen, S. A., Tasyurek, E., Roth, J. E. & Miller, D. A. B. Measurement and modeling of ultrafast carrier dynamics and transport in germanium/silicon-germanium quantum wells**.** OPTICS EXPRESS 18, 25596-25607 (2010 ).

50      Bonfanti, M., et al. Optical transitions in Ge/SiGe multiple quantum wells with Ge-rich barriers. *Physical Review B* **78**, 041407(R) 14276–14287 (2008).

**Acknowledgements**

The authors would like to thank E. Gatti, E. Bonera, A. Marzegalli and F. Montalenti for fruitful discussions. This work was supported by Regione Lombardia through Dote ricercatori, the Swiss Federal program funding Nano-Tera thorough project NEXRAY, and the EC through the GREEN Silicon project (No 257750). Partial support from Cariplo foundation via MANDIS project is also acknowledged.

**Additional information**

Correspondence and requests for materials should be addressed to F.P.

**Competing financial interests:**

The authors declare no competing financial interests.

**Figure captions**

**Figure 1: Sample morphology.** Perspective-view SEM micrographs of the 8 μm Ge-crystals grown on **a** 2×2 μm$^2$, **b** 5×5 μm$^2$ and **c** 9×9 μm$^2$ Si(001) pillars. **d**, Top surface coverage of the (001) facet as a function of the base size of the pristine Si pillars.

**Figure 2: Optical properties of 3D Ge-crystals. a** Room temperature (RT) PL spectra of Ge-crystals grown on Si pillars having a square base of 2×2 μm$^2$ (blue line), 5×5 μm$^2$ (red line), 9×9 μm$^2$ (orange line) and Ge grown on flat Si(001) surface (black line). **b** Comparison of the RT PL spectra of the 3D Ge-crystals (blue curve) and bulk Ge (grey curve). The transitions between the conduction band (CB) minima at the $\Gamma_c$ or the $L_c$ points and the $\Gamma_v$ point of the valence band are indicated.

**Figure 3: Interplay between interband and defect-related emission. a** Contour-plot of the temperature-dependent PL for Ge-crystals grown on 2×2 μm$^2$ Si pillars. Spectra at 6, 75, 150 and 225 K are superimposed as references. **b** Integrated intensity of the interband emission (direct, $\Gamma_c - \Gamma_v$, plus indirect, $L_c - \Gamma_v$) and dislocation band as a function of temperature. The blue solid line is the result of charge-dependent barrier modelling. The black solid line is the fitting curve based on an Arrhenius function with two activation energies (see Supplementary Information). The recombination and trapping regimes are separated by a black, dotted vertical line. The inset is a sketch of the luminescence processes taking place in the Ge samples. Following direct gap excitation, holes thermalize towards the edge of the $\Gamma$ valley in the valence band (VB), whereas electrons can thermalize and recombine radiatively either from the $\Gamma_c$ or $L_c$ valleys in the conduction band (CB). Carriers can leak out from VB and CB into the dislocation energy levels (D) (blue arrows), from which they can recombine radiatively or non-radiatively. Red arrows indicates

that, at high lattice temperatures, backward processes can drain carriers out of the dislocation traps, as well as scatter electrons from the $L_c$ to the $\Gamma_c$ minina.

**Figure 4: Bandgap engineering of 3D Ge-crystals. a** Transmission current for electrons (e, black) and holes (h, red) through a 10 nm-thick $Si_{1-x}Ge_x$ reflector as a function of the Ge content of the alloy layer. **b** Integrated intensity of the interband (IB) emission of Ge-crystals as a function of the temperature for crystals without (No Reflector) and with $Si_{0.25}Ge_{0.75}$ reflectors at 2 (A), 4 (B) and 6 (C) μm from the top surface according to the schemes in the insets. IB data are normalised for each sample to the corresponding dislocation-related (DPL) integrated intensity at 6 K. **c** Calculated number of carriers, N, (orange bars) in Ge-crystals normalized to the value $N_0$ of the crystal with no reflectors. Blue bars are the IB/DPL ratio measured at 6K. Data are reported as a function of $Si_{0.25}Ge_{0.75}$ reflectors.

**Figure 5: Optical properties of the quantum well crystals. a** Perspective scanning electron microscope (SEM) image of Ge/SiGe multiple QW crystals grown on 2-μm wide Si pillars. **b** Top-view SEM micrograph of the QW-crystals. **c** Reciprocal space map of the QW-crystals measured around (004) reflection. The superlattice satellite peaks are numbered and indicated by red and green lines. **d** Low temperature PL for QW-crystals grown on 2×2 μm$^2$ Si pillars (blue curve) and on flat (001)Si substrate (black curve). Quantum-confined transitions related to the direct gap at the Γ point and the indirect gap at the L point of Ge are shown along with the dislocation related emission.

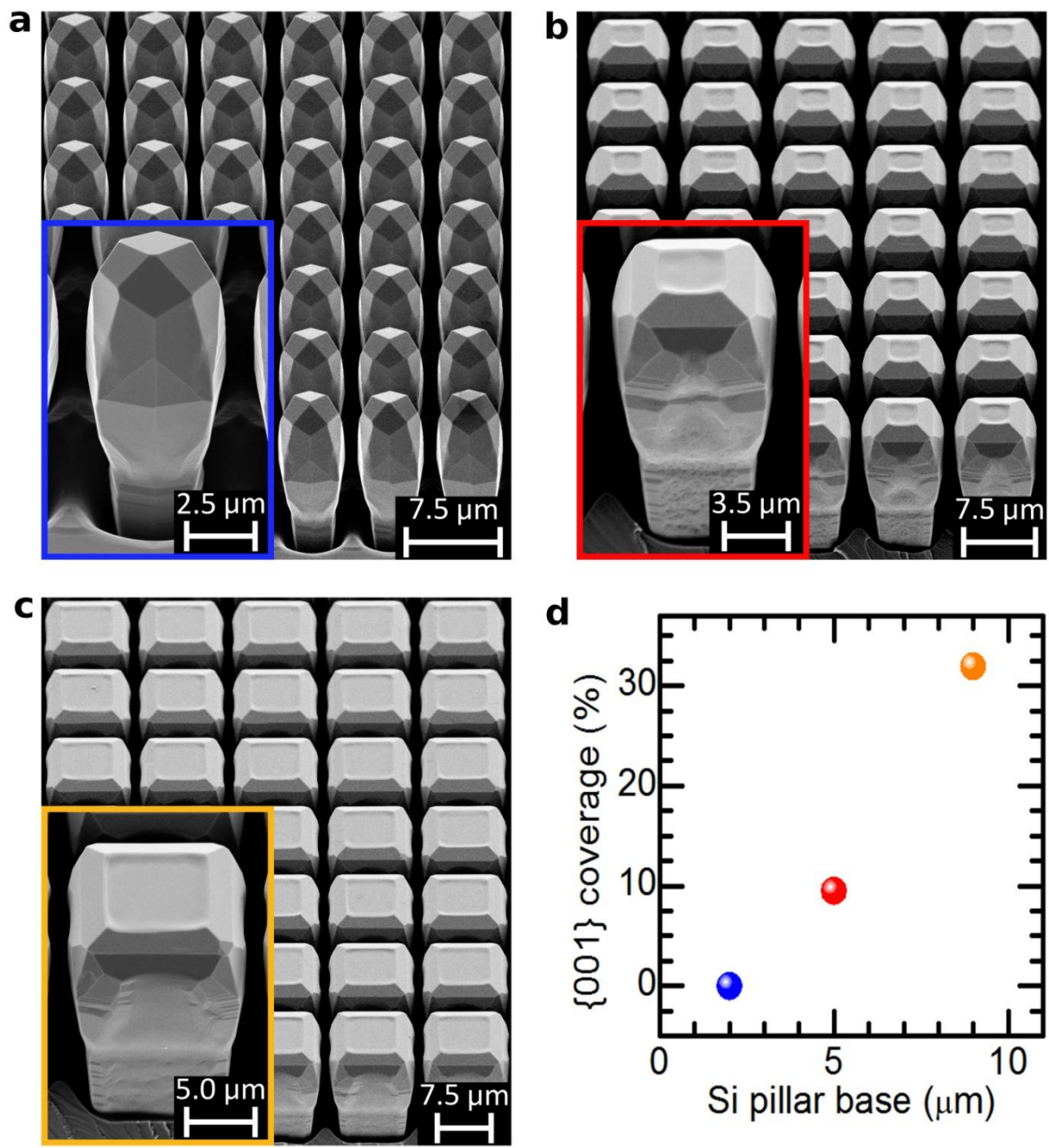

**Figure 1**

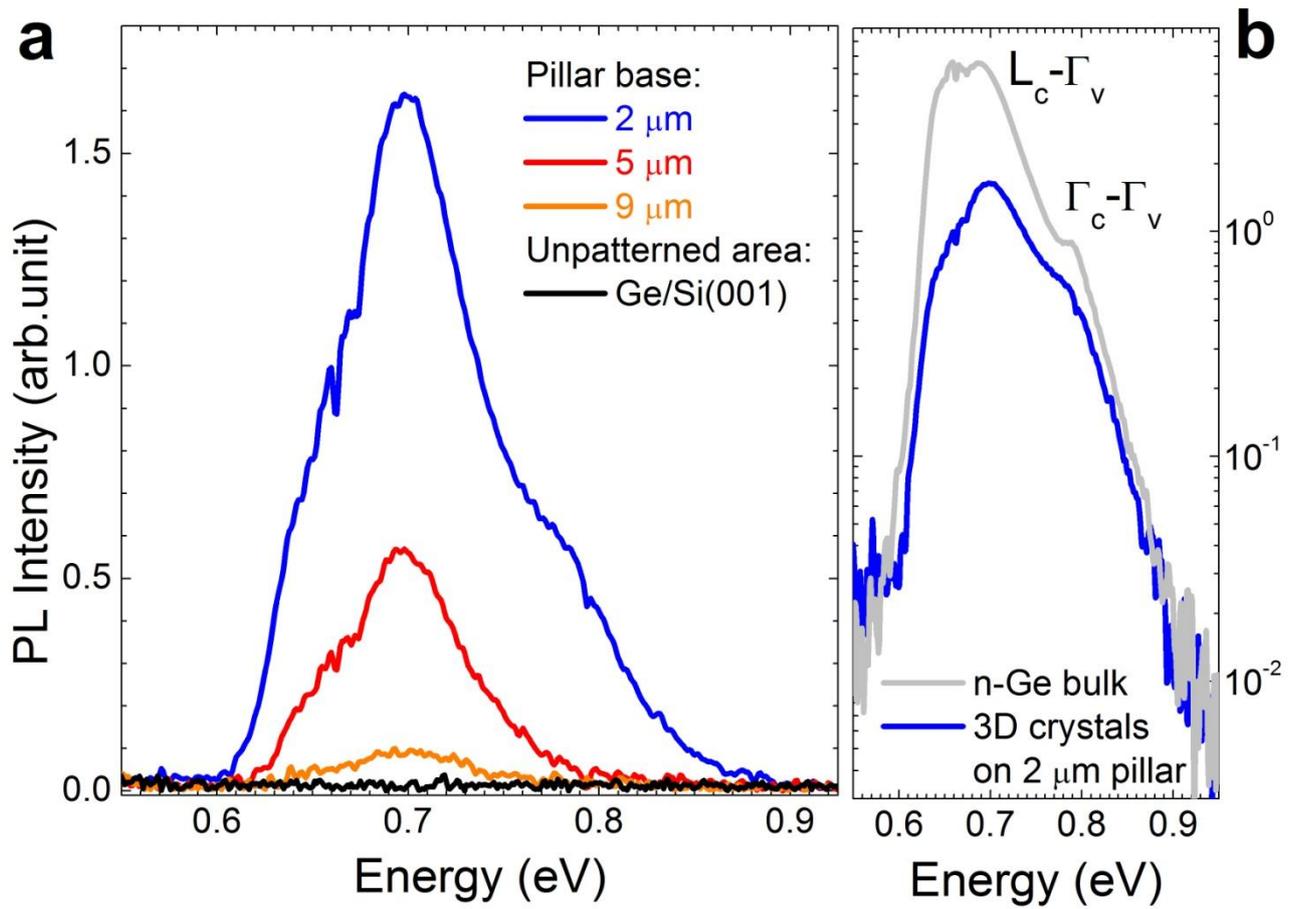

**Figure 2**

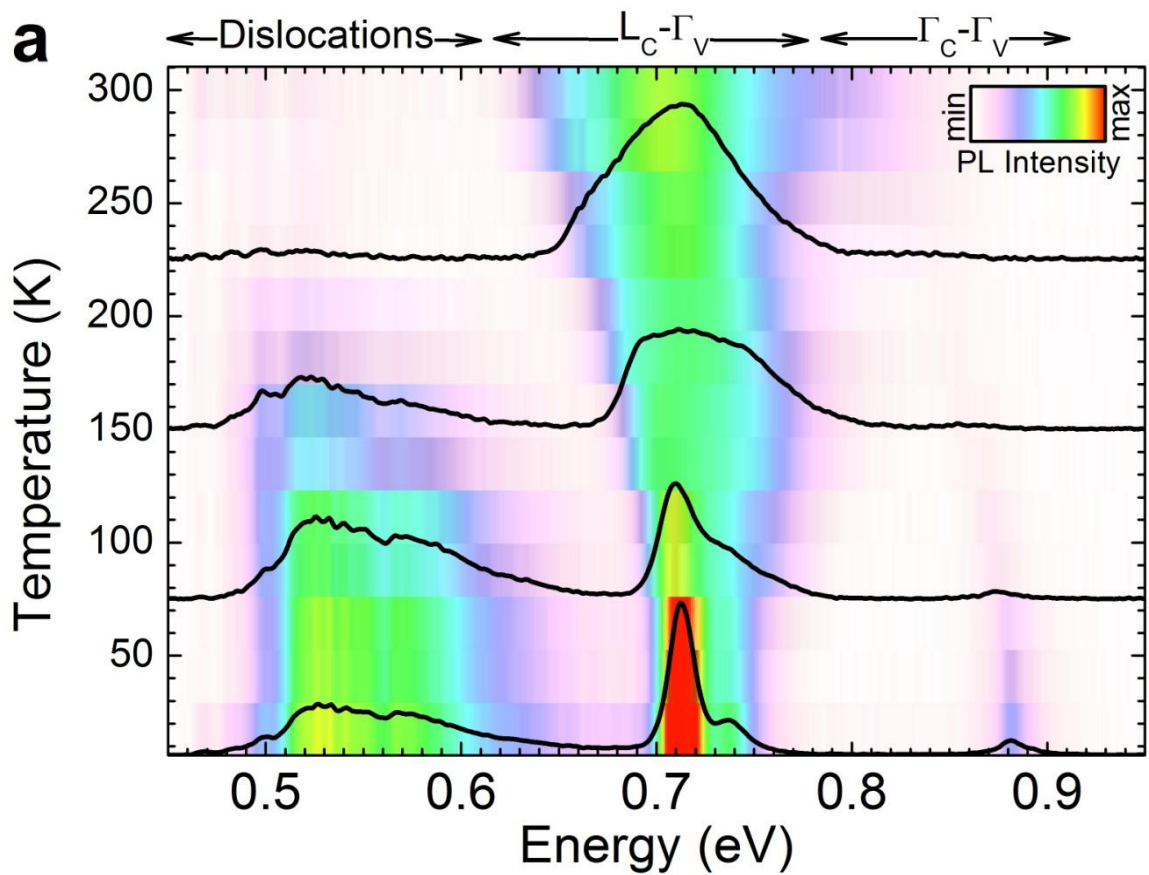

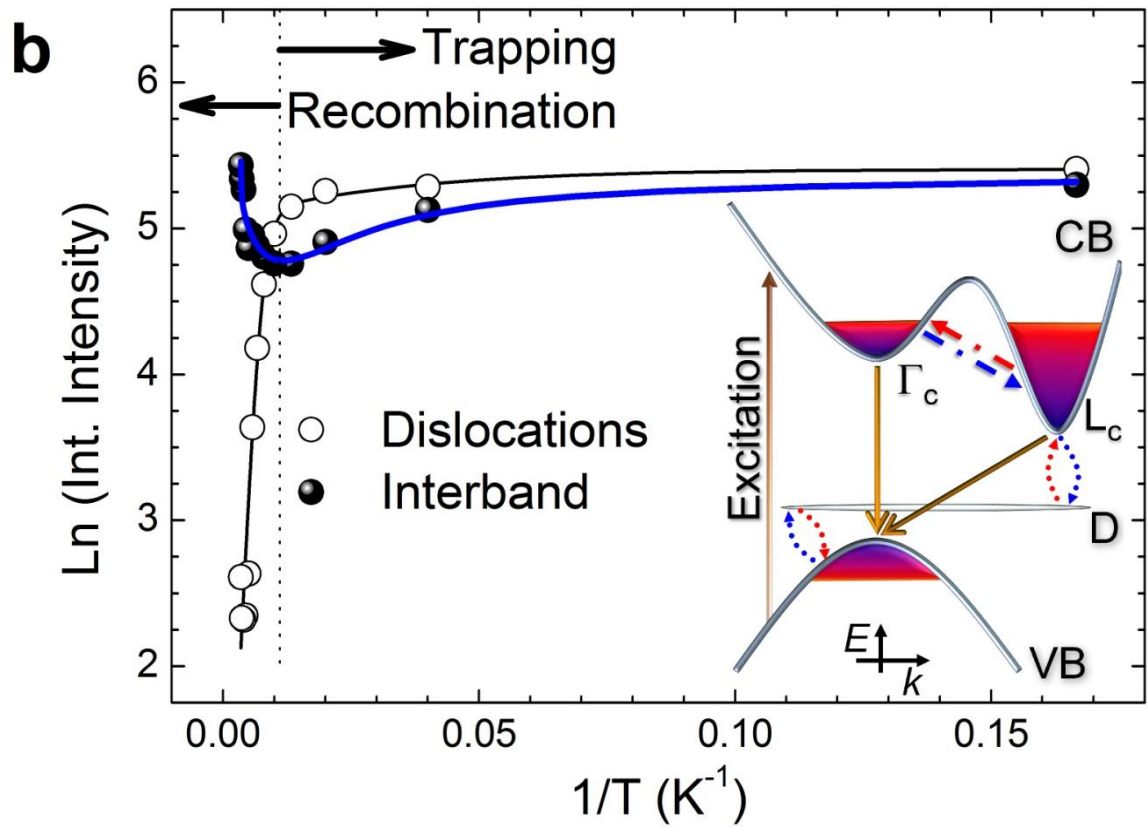

**Figure 3**

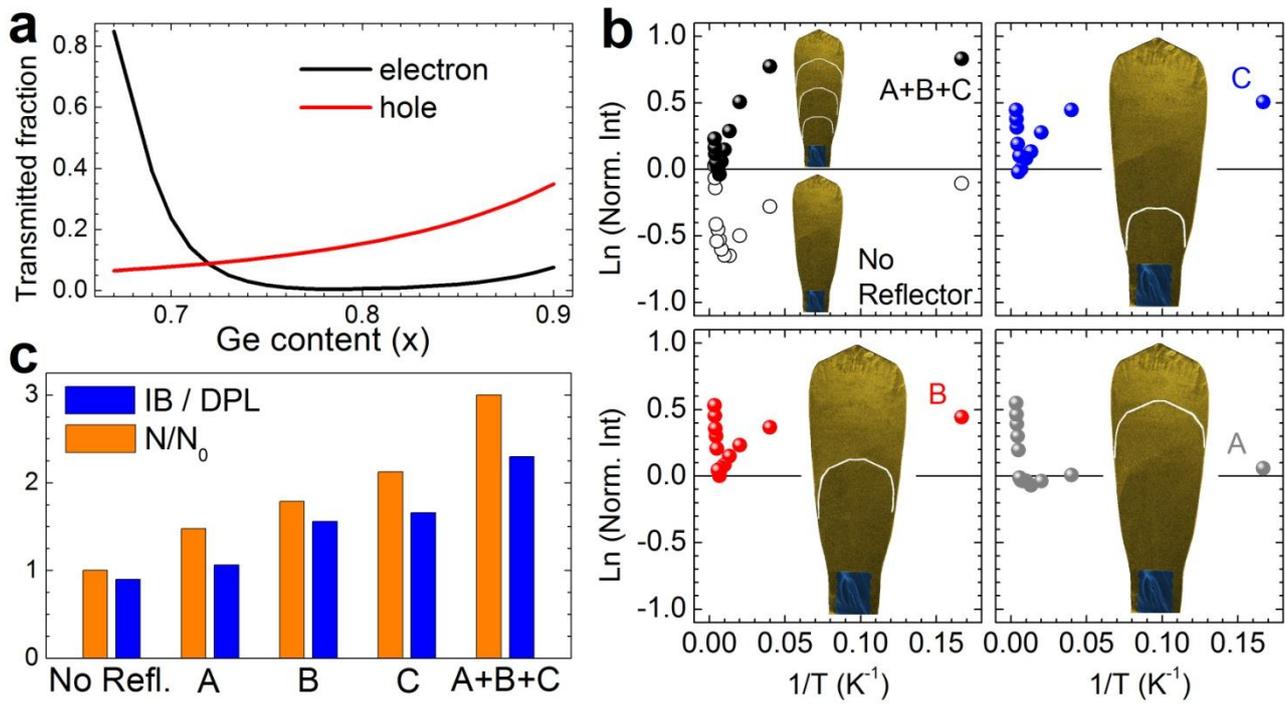

**Figure 4**

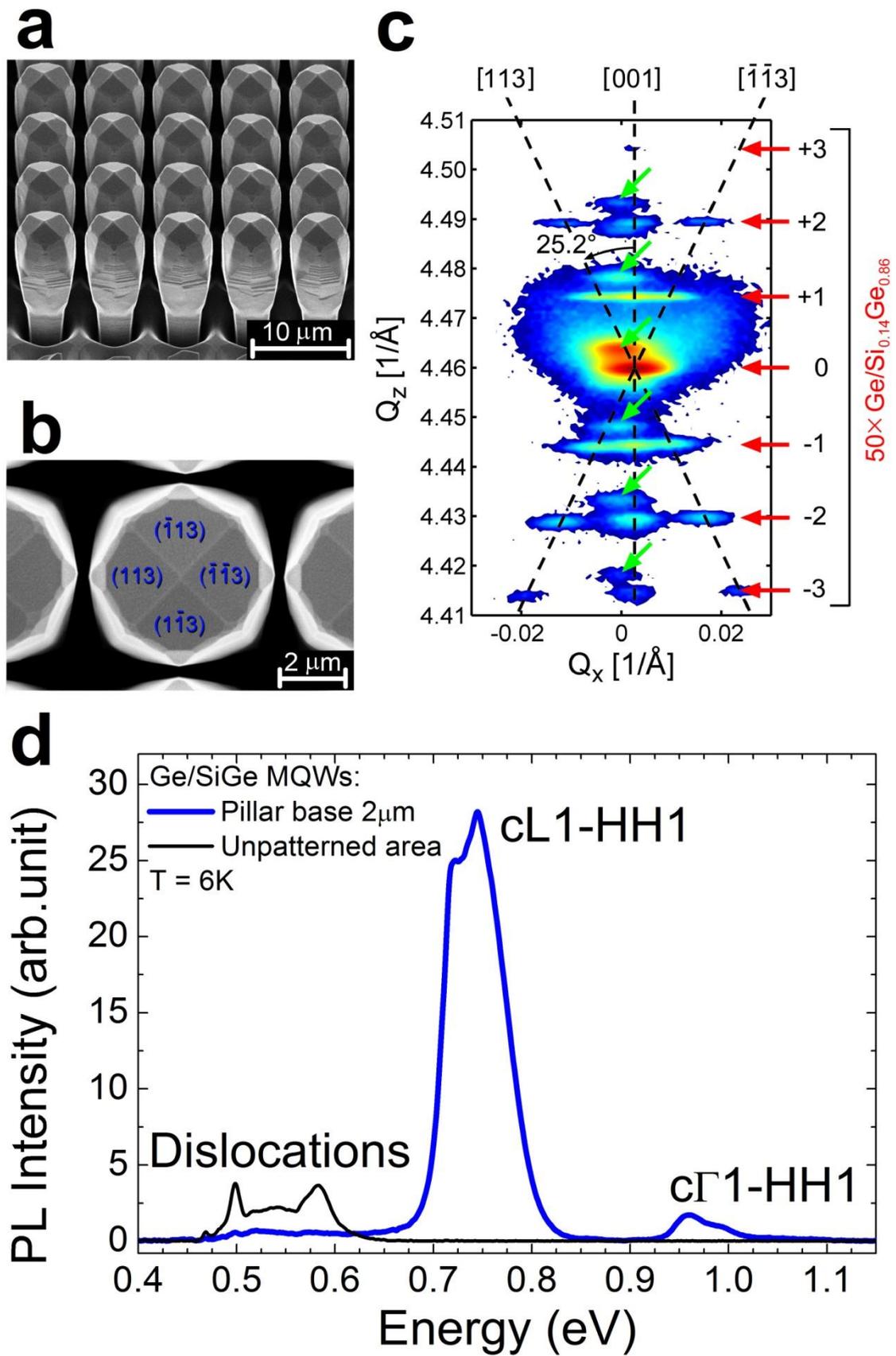

**Figure 5**